\documentclass[12pt,oneside,letterpaper]{article}

\usepackage{amsmath,amssymb,amsfonts}
\usepackage{mathrsfs}
\usepackage{hyperref}
\usepackage{xcolor}

\begin{document}

\begin{center}

~
\vskip5mm

{\large\textbf{An Algebraic Resolution of the Firewall Paradox}}

\vskip8mm

Naman Kumar 

\vskip8mm
{ \it Department of Physics\\
Indian Institute of Technology Gandhinagar\\
Palaj, Gujarat, India, 382355}

\tt{namankumar5954@gmail.com, naman.kumar@iitgn.ac.in}

\end{center}

\vspace{4mm}

\date{\today}
\begin{abstract}
The AMPS firewall argument relies on treating early radiation, late outgoing
Hawking modes, and interior partner modes as approximately independent quantum
subsystems. In diffeomorphism-invariant quantum gravity, however, gravitational
dressing and asymptotic constraints obstruct such a tensor-product
factorization of physical observables. In this essay, we sharpen this
obstruction by formulating subsystem independence directly in
operator-algebraic terms. Using modular theory, half-sided modular inclusions
along null directions, and the sector-wise maximality of the dressed radiation
algebra at future null infinity, we show that---within a fixed asymptotic charge
sector---the algebra associated with the interior Hawking partner cannot form an
independent commuting subalgebra, but must be contained as a (non-commuting) subalgebra of the radiation algebra itself. The subsystem-independence assumption underlying the
AMPS paradox therefore fails, and the entanglement-monogamy step never becomes
applicable. As a result, unitary black hole evaporation and semiclassical
horizon smoothness are compatible in asymptotically flat quantum gravity,
without invoking entanglement islands, replica wormholes, or modifications of
semiclassical horizon physics.
\\
\\
\\
{\centering \emph{This Essay received an honorable mention in the Gravity Research Foundation Awards for Essays on Gravitation 2026}}
\end{abstract}

\section{Introduction}

The firewall paradox of Almheiri, Marolf, Polchinski, and Sully~\cite{Almheiri:2012rt}
(AMPS) exposes a sharp tension between three assumptions that are each
independently well motivated within semiclassical gravity:
(i) \emph{unitary black hole evaporation}, as required by quantum mechanics;
(ii) \emph{semiclassical smoothness at the event horizon}, as implied by the
equivalence principle and the local validity of quantum field theory in curved
spacetime; and (iii) \emph{subsystem independence} of the interior Hawking partner
mode from the early radiation. For an old black hole, the AMPS argument shows
that these assumptions cannot be simultaneously maintained: unitarity requires
strong correlations between late-time Hawking quanta and the early radiation,
while horizon smoothness requires those same quanta to be maximally entangled
with interior partner modes, apparently violating entanglement monogamy.

The first two assumptions are deeply rooted in both experiment and theoretical
consistency. The third, however, rests on a structural premise that is often left
implicit—namely, that the interior partner mode defines an independent quantum
subsystem, described by a Hilbert-space factor that approximately commutes with
the algebra of exterior observables. Such a factorization is natural in
nongravitational quantum field theory, where locality and microcausality
guarantee the existence of independent subsystems. Its validity in
diffeomorphism-invariant quantum gravity, however, is far from obvious.
Gravitational constraints, long-range gauge fields, and the intrinsically
relational nature of physical observables obstruct a naive decomposition of the
physical Hilbert space into interior and exterior tensor factors.

These obstructions have been emphasized in various forms in earlier work,
notably by Raju, Laddha and collaborators~\cite{Papadodimas:2012aq,Papadodimas:2013jku,Laddha:2020kvp,Raju:2021lwh},
who argued that the AMPS paradox arises from importing a nongravitational notion
of subsystem independence into quantum gravity. In this view, the absence of a
tensor-product decomposition—due to gauge constraints and gravitational
dressing—already undermines the assumption that the interior Hawking partner
constitutes an independent system, thereby dissolving the paradox at a
conceptual level.

The present work sharpens this perspective by formulating subsystem
independence directly in operator-algebraic terms. Rather than diagnosing the
problem at the level of Hilbert-space factorization, we characterize
independence by the existence of commuting von Neumann subalgebras of
gauge-invariant observables. Working in asymptotically flat spacetimes, we
explicitly incorporate gravitational dressing and asymptotic constraints and
show that, within a fixed asymptotic charge sector, the algebra associated with
the interior Hawking partner cannot form an independent commuting subalgebra.
Instead, it must be contained as a (non-commuting) subalgebra of the asymptotic radiation algebra.

This algebraic inclusion identifies precisely where the AMPS reasoning fails.
The tripartite tensor-product structure required for the application of
entanglement monogamy simply does not exist in quantum gravity once gauge
invariance and boundary-generated dynamics are properly taken into account.
As a result, the entanglement-monogamy step never applies, and the firewall
conclusion is avoided without sacrificing either unitary evaporation or
semiclassical horizon smoothness.

\section{Radiation algebra at $\mathscr I^+$, Gravitational dressing and interior observables}
We work in an asymptotically flat spacetime and define the radiation algebra
$\mathcal A_R$ as the von Neumann algebra generated by all gauge-invariant
observables localized on future null infinity $\mathscr I^+$. Concretely,
$\mathcal A_R$ contains the algebra generated by the Bondi news operators
$N_{AB}(u,\Omega)$, which encode the hard radiative degrees of freedom, together
with the soft sector describing gravitational memory and zero modes. In addition,
$\mathcal A_R$ includes the generators of asymptotic symmetries, namely the BMS
charges, which act as boundary observables characterizing the global gravitational
state.

The presence of asymptotic charges implies that $\mathcal A_R$ possesses a
nontrivial center, corresponding to superselection sectors labeled by fixed
values of the conserved charges (e.g., total energy--momentum and supermomentum).
To obtain a factor algebra appropriate for discussing quantum correlations, we
restrict attention to a fixed superselection sector $\alpha$ and denote the
resulting von Neumann factor by $\mathcal A_R^{(\alpha)}$. All statements below
are understood to hold within such a fixed sector, where the center has been
quotiented out and standard modular notions apply.\vspace{2mm}

A crucial structural feature of diffeomorphism-invariant theories is the absence
of strictly local gauge-invariant observables. Any operator that purports to
measure a local bulk degree of freedom must be defined relationally and
accompanied by an appropriate gravitational dressing~\cite{Giddings:2007nu,Giddings:2019hjc}. Schematically, a dressed
bulk operator takes the form
\begin{equation}
\Phi(x)=\phi(x)\,\mathcal D_g(x;\Gamma),
\end{equation}
where $\phi(x)$ is a local field operator and $\mathcal D_g(x;\Gamma)$ is a
functional of the gravitational field specifying a relational reference frame or
``dressing'' along a curve $\Gamma$ that extends to asymptotic infinity. The
precise form of the dressing is not essential; what matters is that it necessarily
involves asymptotic gravitational degrees of freedom.

As a consequence, operators associated with the would-be interior Hawking partner
mode—although geometrically localized behind the horizon in a semiclassical
description—cannot be elements of an algebra that is independent of the
asymptotic radiation. Once properly dressed to ensure diffeomorphism invariance,
such operators act nontrivially on the asymptotic gravitational field and must be
represented within the global gauge-invariant algebra in the same superselection
sector $\alpha$. This observation already signals a failure of the naive
interior--exterior tensor factorization assumed in the AMPS argument and motivates
an algebraic treatment in which independence is assessed via commutation and
relative commutants rather than by Hilbert-space decomposition.

\section{Modular structure and null translations}
To make contact between near-horizon quantum fields and the algebra of asymptotic
radiation, we exploit the modular structure of local algebras associated with
null regions. Consider a family of von Neumann algebras
$\{\mathcal M(u)\}_{u\in\mathbb R}$, where $\mathcal M(u)$ is generated by
gauge-invariant observables supported on an outgoing-null region bounded by a
retarded-time cut $u$ in a neighborhood of the horizon. These algebras are
naturally nested along the null direction,
\begin{equation}
\mathcal M(u+\delta)\subset \mathcal M(u), \qquad \delta\ge 0,
\end{equation}
reflecting the fact that later null cuts correspond to smaller causal regions.

Let $\Omega$ denote the semiclassical state describing an evaporating black hole,
assumed to be locally vacuum-like in a neighborhood of the horizon. In particular,
$\Omega$ is cyclic and separating for the algebras $\mathcal M(u)$, so that the
Tomita--Takesaki modular theory applies. We further assume that the above inclusion
is a \emph{half-sided modular inclusion} with respect to $(\mathcal M(u),\Omega)$,
meaning that the modular flow generated by $(\mathcal M(u),\Omega)$ leaves
$\mathcal M(u+\delta)$ invariant for one sign of the modular parameter. This
condition is the natural algebraic analogue of the fact that, near a smooth
horizon, modular flow acts geometrically as a boost or dilation along null
generators.

A central result of modular theory then implies the existence of a strongly
continuous one-parameter unitary semigroup $U(\delta)$ with positive generator,
such that~\cite{Haag:1996hvx,Wiesbrock:1992mg}
\begin{equation}
\mathcal M(u+\delta)
=
U(\delta)\,\mathcal M(u)\,U(\delta)^\ast,
\qquad \delta\ge 0.
\end{equation}
The positivity of the generator ensures that $U(\delta)$ implements genuine
future-directed null translations, in accord with the causal structure of the
spacetime. In this sense, half-sided modular inclusions provide an intrinsic,
state-dependent construction of null translation symmetry directly from the
operator algebra.

By iterating this construction and taking the inductive limit along the outgoing
null direction, the union of the translated near-horizon algebras generates the
full algebra of observables associated with radiation up to a given retarded-time
cut. Identifying this inductive limit with the dressed radiation algebra on
$\mathscr I^+$, and subsequently restricting to a fixed superselection sector
$\alpha$, we recover the factor algebra $\mathcal A_R^{(\alpha)}$. Thus, the
modular structure of near-horizon algebras furnishes a precise algebraic bridge
between semiclassical horizon physics and the asymptotic radiation algebra.

\section{Sector-wise maximality from boundary generation of dynamics}
We now motivate a physically natural completeness condition on the dressed
radiation algebra. Working in a fixed asymptotic charge sector $\alpha$, we claim
that the corresponding factor algebra satisfies
\begin{equation}
\big(\mathcal A_R^{(\alpha)}\big)'=\mathbb C\,\mathbf 1,
\label{eq:max}
\end{equation}
i.e.\ there exists no nontrivial gauge-invariant operator commuting with all
radiation observables. 
Equation~\eqref{eq:max} should be understood as a \emph{sector-wise completeness
condition} for asymptotically flat quantum gravity, analogous to Gauss-law
completeness in gauge theories: once the center generated by asymptotic charges
is fixed, no additional gauge-invariant degrees of freedom remain that commute
with the full algebra of dressed radiation observables.
This condition reflects the fact that, once the center
generated by asymptotic charges is fixed, all physical degrees of freedom are
encoded in asymptotic data.

We now provide a dynamical justification of Eq.~\eqref{eq:max} using the
Hamiltonian structure of gravity. In canonical general relativity, the generator
of an asymptotic symmetry with lapse--shift $\xi=(N,N^i)$ takes the form
\begin{equation}
H[\xi]
=
\int_{\Sigma} d^3x\,
\big(N\,\mathcal H + N^i\,\mathcal H_i\big)
\;+\;
Q_{\partial\Sigma}[\xi],
\label{eq:Hboundary}
\end{equation}
where $\mathcal H\approx 0$ and $\mathcal H_i\approx 0$ are the Hamiltonian and
momentum constraints, and $Q_{\partial\Sigma}[\xi]$ is the Regge--Teitelboim/Wald
boundary term at spatial or null infinity. On the physical Hilbert space,
constraints annihilate states (or vanish weakly), so the action of the generator
reduces to a boundary charge,
\begin{equation}
H[\xi]\;\widehat{=}\;Q_{\partial\Sigma}[\xi]
\qquad \text{on } \mathcal H_{\rm phys}.
\label{eq:HequalsQ}
\end{equation}
In particular, time evolution is generated by the asymptotic energy
(ADM/Bondi charge), and more generally by the full set of asymptotic symmetry
charges.

This ``gravitational Gauss law'' implies that any gauge-invariant bulk observable
must be gravitationally dressed so as to transform appropriately under
$Q_{\partial\Sigma}[\xi]$~\cite{Regge:1974zd,Wald:1999wa}. Equivalently, there can be no strictly localized
gauge-invariant subalgebra that is dynamically decoupled from asymptotic data.

Let $\mathcal A_R$ denote the dressed algebra on $\mathscr I^+$ generated by the
hard radiative modes together with the soft sector and asymptotic charges. Since
the charges generate a nontrivial center, we fix a superselection sector $\alpha$
and work with the corresponding factor algebra $\mathcal A_R^{(\alpha)}$.
Consider any bounded operator $X$ acting on $\mathcal H_{\rm phys}^{(\alpha)}$
that commutes with all radiation observables, i.e.\
$X\in(\mathcal A_R^{(\alpha)})'$. In particular, $X$ commutes with the boundary
Hamiltonian and hence with the full unitary time evolution,
\begin{equation}
[X,\,Q_{\partial\Sigma}[t]] = 0
\quad\Longrightarrow\quad
[X,\,U(t)] = 0 .
\label{eq:XcommutesU}
\end{equation}
Moreover, $X$ commutes with the complete algebra of radiative observables on
$\mathscr I^+$, including the Bondi news and memory operators.

Assuming that $\mathcal A_R^{(\alpha)}$ acts cyclically on
$\mathcal H_{\rm phys}^{(\alpha)}$—that is, the dressed radiation algebra
generates the physical state space in the fixed sector—it follows that any
operator commuting with all of $\mathcal A_R^{(\alpha)}$ must act as a multiple of
the identity. This establishes the sector-wise maximality condition
Eq.~\eqref{eq:max}. Cyclicity is physically equivalent to the statement that the complete set of
gravitationally dressed radiative observables on $\mathscr I^+$ suffices to
generate all physically accessible states in the fixed asymptotic charge sector, as expected
for unitary black hole evaporation~\cite{Marolf:2008mf}.
\vspace{2mm}

The essential structural input is that in gravity the generators
of dynamics and asymptotic symmetries reduce to boundary charges once constraints
are imposed, cf.\ Eqs.~\eqref{eq:Hboundary}--\eqref{eq:HequalsQ}. The inclusion of
the soft sector is crucial: without the BMS and memory degrees of freedom, the
asymptotic algebra would fail to be closed under gauge invariance and would not
admit the sector-wise factorization required for Eq.~\eqref{eq:max}.

\subsection{Relation to islands, long-range gravity, and compactly
supported operators}

The obstruction described above is closely related to the recent debate on
whether entanglement islands are compatible with long-range gravity.  The
island formula and replica-wormhole calculations provide a powerful
semiclassical mechanism for recovering the Page curve.  In that language,
the interior island is encoded in the Hawking radiation rather than being an
independent microscopic system
\cite{Almheiri:2019psf,Almheiri:2019hni,Penington:2019kki,
Almheiri:2019yqk,Almheiri:2020cfm}.  This interpretation is precisely what
makes the question subtle in a theory with a massless graviton: if the island
degrees of freedom are to be treated as an entanglement wedge disconnected
from the asymptotic region, then the corresponding operator algebra should
commute with the algebra of the complement.

Geng and Karch emphasized that many controlled higher-dimensional island
constructions effectively involve a massive graviton, and that the island
contribution disappears in the zero-mass limit in their model
\cite{Geng:2020fxl}.  More generally, Geng et al. argued that an island in a
standard long-range gravitational theory is in tension with the gravitational
Gauss law: an excitation localized in the island changes the asymptotic
gravitational field and can therefore be detected from outside the island,
whereas ordinary entanglement-wedge complementarity would require the
island algebra to commute with the algebra of the complement
\cite{Geng:2021hlu}.  This line of reasoning suggests that massless gravity
does not naturally admit a sharply independent island algebra in the same
way as a nongravitating bath or a massive-gravity setup.

Recently, Ref.~\cite{Antonini:2025rkh} has challenged the strongest
form of this conclusion.  It argues that, on sufficiently generic backgrounds
with no exact isometries, gauge-invariant operators with compact support can
be constructed to all orders in perturbation theory.  Our claim is compatible
with this perturbative statement.  The present essay does not deny the
emergence of local bulk operators in a semiclassical code subspace.  Rather,
it distinguishes such perturbative or code-subspace locality from the exact
operator-algebraic independence required by the AMPS argument.

The AMPS monogamy step requires more than the existence of approximately
local semiclassical representatives.  It requires an interior partner algebra
which is an independent commuting factor, namely
\begin{equation}
        \mathcal A_C^{(\alpha)} \subset
        \bigl(\mathcal A_R^{(\alpha)}\bigr)' .
\end{equation}
By contrast, perturbatively compact operators need not define an exact
nonperturbative relative commutant of the full radiation algebra.  They may
realize emergent locality within a restricted code subspace, while still
failing to produce a separate factor in the exact gravitational Hilbert
space.  This distinction is crucial because entanglement monogamy is a
statement about independent tensor factors, or equivalently about commuting
operator algebras, not merely about the existence of useful semiclassical
operator representatives.

In fact, when island reconstruction works, it supports the same algebraic
lesson emphasized here: the would-be island degrees of freedom are not a
third microscopic subsystem independent of the radiation, but are represented
in the radiation algebra itself.  The present essay isolates this common
algebraic core directly, without relying on the replica derivation of the
island formula.  The interior partner should therefore not be placed in the
relative commutant of the radiation algebra.  In the exact gravitational
algebra relevant for the AMPS argument, it is encoded in
$\mathcal A_R^{(\alpha)}$, not in
$\bigl(\mathcal A_R^{(\alpha)}\bigr)'$.

\section{Inclusion of the interior partner algebra and implications for the firewall paradox}
Let $\mathcal A_C^{(\alpha)}$ denote the von Neumann algebra generated by the
gravitationally dressed observables associated with the interior Hawking partner
mode, defined within the same asymptotic charge sector $\alpha$ as the radiation
algebra. The standard AMPS reasoning assumes that the interior partner defines an
\emph{independent} quantum subsystem, so that its observables satisfy
microcausality with respect to the asymptotic radiation. In algebraic terms, this
assumption requires that the interior observables commute with all radiation
observables,
\begin{equation}
[\mathcal A_C^{(\alpha)},\,\mathcal A_R^{(\alpha)}]=0,
\end{equation}
which is equivalent to the statement that the interior algebra is contained in
the commutant of the radiation algebra,
\begin{equation}
\mathcal A_C^{(\alpha)} \subseteq (\mathcal A_R^{(\alpha)})'.
\label{eq:ACcommutant}
\end{equation}

However, the sector-wise maximality condition established in
Eq.~\eqref{eq:max} implies that the commutant of the dressed radiation algebra is
trivial,
\begin{equation}
(\mathcal A_R^{(\alpha)})' = \mathbb C\,\mathbf 1.
\end{equation}
This result expresses the fact that, once gravitational dressing and asymptotic
constraints are properly taken into account, there exists no nontrivial
gauge-invariant algebra that is independent of the radiation observables within
a fixed charge sector. Since the interior Hawking partner mode carries
nontrivial physical information—such as energy, entropy, and correlations with
the outgoing radiation—its observables cannot be proportional to the identity.
Therefore, the assumption of subsystem independence encoded in
Eq.~\eqref{eq:ACcommutant} is incompatible with the algebraic structure of
asymptotically flat quantum gravity.

The resolution is that the interior partner observables do not form a disjoint
tensor factor, nor a commuting subalgebra, but are instead contained as a (non-commuting) subalgebra of the radiation algebra. Algebraically, one must have the strict inclusion
\begin{equation}
\mathcal A_C^{(\alpha)} \subseteq \mathcal A_R^{(\alpha)}
\label{eq:inclusion}
\end{equation}
indicating that the interior degrees of freedom are encoded—albeit in a highly
nonlocal and state-dependent manner—within the operator algebra of the asymptotic
radiation.\vspace{2mm}

The AMPS firewall argument relies essentially on the monogamy of entanglement,
which is a statement about correlations in tripartite quantum systems admitting
a tensor-product Hilbert-space structure
$\mathcal H_R \otimes \mathcal H_B \otimes \mathcal H_C$. In the AMPS reasoning,
the early radiation $R$, the late outgoing Hawking mode $B$, and the interior
partner $C$ are treated as three approximately independent subsystems, so that
maximal entanglement between $B$ and $R$ is taken to exclude simultaneous maximal
entanglement between $B$ and $C$.

The algebraic relation~\eqref{eq:inclusion} invalidates this premise. Since the
interior algebra $\mathcal A_C^{(\alpha)}$ is embedded within the radiation
algebra $\mathcal A_R^{(\alpha)}$, the correlations between $B$ and $C$ are not
correlations between independent subsystems, but rather correlations internal to
the radiation sector itself. As a result, the tripartite tensor-product structure
required for the application of entanglement monogamy is absent, and there is no
third independent system with which the late radiation could violate it.

It follows that unitary evaporation and semiclassical horizon smoothness are not
in tension in asymptotically flat quantum gravity. The firewall conclusion is
therefore not forced by the underlying physics, but arises from an unjustified
assumption of subsystem independence that fails once the operator-algebraic
structure of gravitational observables is treated correctly.

\section{Conclusion}

In this essay, we have shown that the firewall paradox can be resolved within
asymptotically flat quantum gravity once the notion of subsystem independence
is formulated in a manner consistent with gauge invariance and asymptotic
constraints. Using only structural ingredients intrinsic to gravity—the
operator-algebraic formulation of observables, gravitational dressing, modular
theory along null directions, and the sector-wise maximality of the dressed
radiation algebra—we have demonstrated that the interior Hawking partner does
not define an independent quantum subsystem. In a fixed asymptotic charge
sector, any gauge-invariant operator associated with the interior must act
within the same von Neumann algebra as the asymptotic radiation, and cannot
belong to a commuting factor.

The central result is the algebraic inclusion
$\mathcal A_C^{(\alpha)} \subseteq \mathcal A_R^{(\alpha)}$, which directly
invalidates the assumption that the interior partner may be treated as a
separate system entangled with the outgoing Hawking mode. Once gravitational
dressing and asymptotic constraints are properly incorporated, the interior
algebra cannot commute with the full radiation algebra without becoming
trivial. The subsystem independence required by the AMPS argument is therefore
incompatible with the operator-algebraic structure of asymptotically flat
quantum gravity.

From this perspective, the firewall paradox does not signal a fundamental
conflict between unitary evaporation and semiclassical horizon smoothness.
Rather, it arises from importing a nongravitational notion of Hilbert-space
factorization into a theory in which gauge constraints and boundary-generated
dynamics preclude such a decomposition. In the absence of a tensor-product
structure separating early radiation, late outgoing modes, and interior
partners, the entanglement-monogamy step at the core of the AMPS argument never
becomes applicable.

Our analysis is purely algebraic and does not rely on replica wormholes,
entanglement islands, or modifications of semiclassical horizon physics.
Instead, it highlights a basic structural distinction between gravity and
ordinary quantum field theory: in gravity, physical degrees of freedom are
intrinsically relational and encoded in asymptotic data, so subsystem structure
must be assessed at the level of operator algebras rather than Hilbert-space
tensor factors. When this distinction is respected, black hole evaporation can
remain unitary while preserving a smooth horizon—not because new dynamics are
introduced, but because gravity itself does not admit the subsystem structure
assumed by the firewall argument.

\bibliography{AMPS}
\bibliographystyle{plain}

\end{document}